\documentclass[draft,linenumbers]{agujournal2018}
\usepackage{apacite}
\usepackage{amsmath}
\usepackage{url} 
\usepackage{float}
\usepackage[caption = false]{subfig}
\usepackage{graphicx}
\draftfalse

\journalname{JGR-Space Physics}

\begin{document}

\title{Evaluation of plasma properties from chorus waves observed at the generation region }

\authors{Lilla Juh\'asz\affil{1},Yoshiharu Omura\affil{2}, J\'anos Lichtenberger\affil{1,3}, Reinhard H. Friedel\affil{4}}

\affiliation{1}{Department of Geophysics and Space Science, E\"otv\"os University, Budapest Hungary}
\affiliation{2}{Research Institute of Sustainable Humanosphere, Kyoto University, Kyoto, Japan}
\affiliation{3}{Research Center for Astronomy and Earth Sciences, Hungarian Academy of Sciences, Sopron, Hungary}
\affiliation{4}{Space Science and Applications Group, Los Alamos National Laboratory, Los Alamos, NM, USA}

\correspondingauthor{Lilla Juhász}{lilla@sas.elte.hu}

\begin{keypoints}
\item Linear growth of whistler-mode waves forms triggering waves for nonlinear chorus emissions. 
\item We have developed an inversion method to derive energetic electron properties from observation of chorus emissions.
\item The particle properties derived from chorus-inversion agree with energetic particle measurements.
\end{keypoints}

\begin{abstract}

In this study we present an inversion method which provides thermal plasma population parameters from characteristics of chorus emissions only. Our ultimate goal is to apply this method to ground based data in order to derive the lower energy boundary condition for many radiation belt models. The first step is to test the chorus-inversion method on in-situ data of Van Allen Probes in the generation region.
Density and thermal velocity of energetic electrons (few keV - 100 keV) are derived from frequency sweep rate and starting frequencies of chorus emissions through analysis of wave data from Electric and Magnetic Field Instrument Suite and Integrated Science (EMFISIS) onboard the Van Allen Probes. Nonlinear wave growth theory of \citet{omura2011triggering} serves as the basis for our inversion method, assuming that the triggering wave is originated by the linear cyclotron instability. We present sixteen, consecutive rising-tone emissions recorded in the generation region between 11-12UT on 14 November 2012. The results of the inversion are compared with density and thermal velocities (parallel and perpendicular) of energetic electrons derived from unidirectional flux data of Helium Oxygen Proton Electron (HOPE) instrument, showing a good agreement: the normalized root-mean-square deviation between the measured and predicted values are $\sim13\%, \sim6\%$, and $\sim10\%$, respectively. We found that the theoretical amplitudes are consistent with the measured ones. The relation between linear and nonlinear wave growth agrees with our basic assumption, namely, linear growth is a preceding process of nonlinear wave growth. We analyze electron distributions at the relativistic resonant energy ranges. 
\end{abstract}

\section{Introduction}
\label{intro}

In recent years' NASA missions such as Time History of Events and Macroscale Interactions during Substorms (THEMIS), Van Allen Probes (RBSP) and  Magnetospheric Multiscale (MMS) advanced our understanding of the complex inter-connections of geospace environment because of the availability of in-situ data. Some of these in-situ data are the boundary conditions and parametric input to many space environment models, and are critical to enable accurate now-casts and forecast. However, a trusted operational system would rely on continuous and long-running measurements of them. A solution for that need can be ground-based measurements of key parameter inputs. The PLASMON project (PLASmasphere MONitoring, an FP7-SPACE-2010-1 Collaborative Project) is an outstanding example for efforts to produce important key parameters, like plasmasphere densities, with the use of ground-based whistler measurements [\cite{lichtenberger2013plasmasphere}]. As part of PLASMON, the global AWDANet network [\cite{lichtenberger2008automatic}; \cite{lichtenberger2009new}] - consisting  of 28 VLF receiver stations -, can be extended with the capability of recording whistler mode chorus emissions at stations with magnetic footprint $L > 4 (3)$. In particular, we will show in this paper how rising tone chorus emissions can be used as a proxy to estimate the in-situ thermal plasma conditions, which form the low-energy boundary condition of many of our current state-of-the-art radiation belt and ring current models. 

Coherent chorus emissions are typically observed as rising/falling tones in the frequency range of $0.1 f_{ce} < f < 0.8 f_{ce}$ with discontinuity at $0.5 f_{ce}$, where $f_{ce}$ is the electron gyrofrequency [\cite{JGR:JGR7608}; \cite{KOONS19901335}; \cite{santolik2003spatio}; \cite{SAZHIN1992681} ]. These emissions are typically excited during geomagnetic storms close to the magnetic equator in low-density plasmas near outside the plasmapause. Chorus emissions are known to be generated via wave-particle interactions with an anisotropic distribution of energetic electrons (few keV- 100 keV) injected from the plasmasheet  [\cite{kennel1966limit} ; \cite{anderson1977vlf} ; \cite{ledocq1998chorus}; \cite{meredith2001substorm}; \cite{omura2009nonlinear}; \cite{santolik2010wave‐particle}; \cite{li2013characteristics}; \cite{JGRA:JGRA51364}].  Anisotropic angular distributions of substorm injected energetic electrons (also called source population [\cite{JGRA:JGRA51985}]) are able to provide free-energy for chorus wave excitation [\cite{thorne2013rapid}, and references therein] and cause isotropic pitch angle distribution (PAD) in the energy range of the interacting particles. Attention of radiation belt modelers recently turned to whistler mode chorus waves due to its role in both accelerating electrons to MeV energies in the Earth's outer radiation belt [\cite{horne1998potential}; \cite{summers1998relativistic}; \cite{summers2002model}; \cite{reeves2013electron}; \cite{thorne2013rapid}; \cite{li2014radiation}] and in pitch angle scattering of electrons into the atmospheric loss cone [\cite{lorentzen2001observations}; \cite{o2004quantification}; \cite{thorne2005timescale}; \cite{hikishima2010microburst}]. The generation of chorus emissions is known to be driven by electron cyclotron resonance  [\cite{kennel1966limit}; \cite{kennel1967unstable}; \cite{tsurutani1974postmidnight}; \cite{nunn1997numerical} ; \cite{chum2007chorus}; \cite{katoh2007computer}, \cite{katoh2007relativistic};\cite{omura2008theory}].\\
\citet{omura2008theory} and \citet{omura2011triggering} proposed a nonlinear wave growth theory for chorus wave generation. They assumed that linear instability excites a coherent whistler mode wave which triggers the non-linear process. They found a relationship between measurable characteristics (frequency sweep rate $\partial \omega/\partial t$, optimum wave amplitude $\Omega_{w0}$, threshold amplitude $\Omega_{th}$) of rising-tone emissions and the distribution function of energetic electrons (number density $N_h$, parallel and perpendicular thermal velocity,$V_{t||}$ and $V_{t\perp}$, respectively) participating in wave-particle interaction. Their theory reveals the amplitude dependency of frequency sweep rate of chorus emissions at the generation region close to the magnetic equator. During quasi-parallel propagation away from the magnetic equator, wave amplitude of chorus emissions undergo a convective growth due to the gradient of the magnetic field, but $\partial \omega/\partial t$ is affected only by cold plasma dispersion. During its slightly oblique propagation away from the equator, the gap at $0.5 f_{ce}$ is formed by nonlinear wave damping via Landau resonance \cite{hsieh2018}.\\
The above mentioned features of the theory led the AWDANet Team to start to develop a method to derive density and thermal velocities of energetic electrons (source population) from chorus emissions recorded on the ground after that they were projected from the ground to the equatorial generation region by a propagation model. When we developed our chorus-inversion method to monitor the equatorial source population, we took into account that the following data are available on AWDANet stations: 1) electromagnetic wave recordings (fs = 20 kHz) 2) equatorial electron plasma number density from PLASMON and 3) electron gyrofrequency obtained from a chosen geomagnetic field model via the station's L value. The 2) and 3) points assume that chorus emissions propagate quasi-parallel to the magnetic field.\\
The main objective of this study is to apply and validate the  chorus-inversion method. The theoretical background of chorus-inversion is described in Section \ref{theo_disc}. In the third section, we present the results of our method on 16 chorus emissions selected from EMFISIS data of Van Allen Probes spacecraft A. Then, we validate the results with simultaneously measured HOPE data from the same spacecraft and analyze the theoretical amplitudes and growth rates. To support the validation process, we  also analyzed the changes of total electron flux and thermal anisotropies from HOPE and Magnetic Electron Ion Spectrometer (MagEIS) instruments. Section \ref{sum_con} gives summary and conclusion.

\section{Determination of thermal velocity and density of energetic electrons}\label{theo_disc}

\begin{figure}[h]
\centering
\includegraphics[width=35pc]{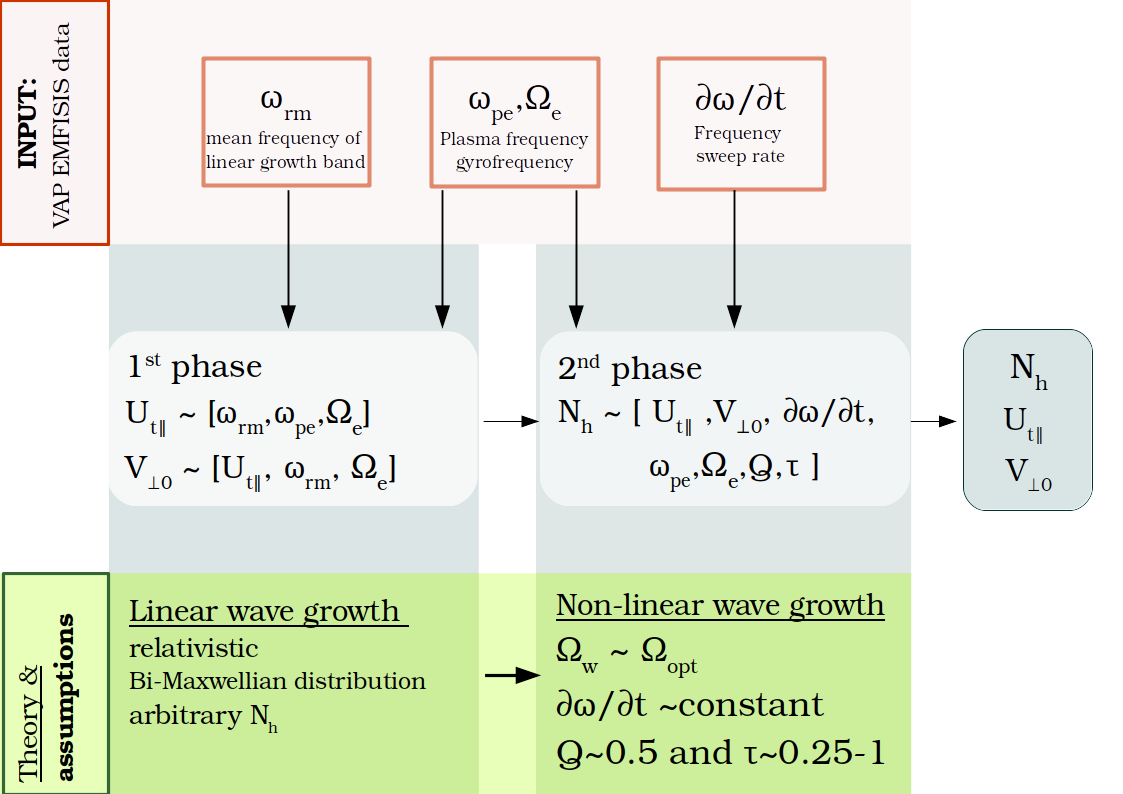}
\caption{Chorus-inversion method: Inputs are from EMFISIS wave measurements (red boxes) only. As the first step, thermal momentum $U_{t||}$ and average perpendicular velocity $V_{\perp 0}$ are calculated assuming that linear wave growth is the initial phase of chorus generation. The second phase is governed by non-linear wave growth. Here, we replace the wave amplitude $\Omega_w$ with the optimum amplitude $\Omega_{opt}$ in order to obtain $N_h$. For the calculation of $N_h$, we use the output of the first phase, $U_{t||}$ and $V_{\perp 0}$. At the end of the process, we obtain the bi-Maxwellian function parameters of energetic electrons responsible for chorus emission generation. In the green boxes we note some important assumptions.}
\label{process}
\end{figure}

The inversion method consists of two phases (Figure \ref{process}). First we estimate the parallel and minimum perpendicular thermal velocity of the source population using the relativistic solution of electromagnetic R-mode wave instability of \citep{doi:10.1063/1.872932} ($1^{st}$ phase blue box in Fig.\ref{process}). Using these thermal velocities, a direct estimation of $N_h/N_c$ is obtained from the frequency sweep rate of a chorus emission using nonlinear wave growth theory ($2^{nd}$ phase blue box). For this study, the inputs are gyrofrequency $\Omega_e$, plasma frequency $\omega_{pe}$, frequency sweep rate of an individual chorus emission $\partial \omega / \partial t$ and the mean frequency of the assumed band of linear growth $\omega_{rm}$, all from EMFISIS measurements (red boxes on Fig.\ref{process}). More about assumptions (green boxes in Fig.\ref{process}) is in the descriptions of the theories mentioned above.

\subsection*{Relativistic linear growth-rate of R-mode plasma waves}

A band of whistler-mode waves is usually present at or below the starting frequency of chorus emissions and acts as a triggering wave for nonlinear wave growth mechanism. This band is assumed to be generated due to relativistic whistler-mode instability that is driven by temperature anisotropy of the source population, $A^M = T_{\perp}/T_{||} -1 = V_{t\perp}^2/V_{t||}^2 -1$ in the case of bi-Maxwellian distribution function. The instability of electromagnetic R-mode waves in a relativistic plasma was studied by  \citet{doi:10.1063/1.872932}. They expressed the linear growth rate as:\\

\begin{equation}\label{lin_inst_growth}
\omega_i = \frac{\pi \omega_{pe}^2 \eta_{rel}}{[2\omega_r + \omega_{pe}^2 |\Omega_e|/(\omega_r - |\Omega_e|)^2]} \{A_{rel} - A_c\},
\end{equation}
where $\eta_{rel}$ is the fraction of the relativistic particle distribution near resonance, which is proportional to the ratio of hot and cold electron density, $N_h/N_c <<$1. 
{$A_{rel}$ is the relativistic pitch-angle anisotropy of the resonant particles, which in the non-relativistic limit is equal to $A^M$. The critical anisotropy is
\begin{equation}\label{AC}
A_c = \frac{1}{\Omega_e/\omega_r -1}.
\end{equation}

In their paper, \citet{doi:10.1063/1.872932} evaluated the linear wave growth rate as a function of frequency $\omega_r$, by numerical integration along the resonance ellipse for different distribution functions, and studied the effects of key parameter changes. In the case of bi-Maxwellian distribution, they found that the variation of $N_h/N_c$ only affects the magnitude of the growth rate. Similarly, the increase of $A_{rel}$ is followed by increasing growth rate, in addition, the frequency range of the instability is slightly spreading. Another important key parameter is the ratio of electron plasma and gyrofrequency $\omega_{pe}/\Omega_e$: decreasing $\omega_{pe}/\Omega_e$ shifts the maximum growth rate to higher frequencies. Likewise, decreasing the hot electron temperature ($U_{t||}$), increases the frequency of the maximum growth rate, also thins the unstable frequency range.\\
We assume that the linear growth rate takes the maximum value at the mean frequency of whistler-instability's wave band $\omega_{rm}$, that frequency is only determined by $\omega_{pe}/\Omega_e$ and $U_{t||}$. In the chorus-inversion $\omega_{pe}/\Omega_e$ and $N_c$ is known, therefore those $U_{t||}$ that produces the maximum linear growth rate of the whistler-mode instability at $\omega_{rm}$ can be the  estimate for initial parallel thermal momentum of source population. Moreover, the minimum resonant anisotropy required for instability, $A_c$, provides the minimum value of $V_{t\perp}$. At this stage of the chorus-inversion, we use an arbitrary $N_h$, because it does not affect the frequency of the maximum value. $N_h$ is calculated in the second step of chorus-inversion method employing the nonlinear wave growth theory.  

\subsection*{Nonlinear wave growth theory}

Linear wave growth induces the initial amplitudes of emissions followed by nonlinear wave growth (\citet{omura2008theory} and \citet{omura2011triggering}) which is responsible for growing amplitude and rising frequency of chorus emissions assuming parallel propagation at the generation region. \citet{omura2009nonlinear} proposed that the formation of gap between upper- and lower-band is due to the nonlinear damping mechanism caused by slightly oblique propagation away from the equator.
The frequency sweep rate of chorus emission is obtained from the inhomogeneity ratio of the relativistic second-order resonance condition at the magnetic equator,

\begin{equation}\label{freqsw}
\frac{\partial \tilde{\omega}}{\partial t} = \frac{0.4 s_0 \omega}{s_1} \tilde{\Omega_{w}},
\end{equation}
where $s_0=\tilde{V}_{\perp 0}\chi/\xi $, $s_1=\gamma(1-\tilde{V}_R/\tilde{V}_g)^2$, $\tilde{\Omega}_{w} = eB_{w}/(m_0 \Omega_{e0})$ and $\tilde{\omega} = \omega/\Omega_{e0}$ is the normalized frequency. $B_w$ is the wave magnetic field, $ \chi^2=(1+\xi^2)^{-1}$ and $ \xi^2=\omega(\Omega_e-\omega)/\omega_{pe}^2$, $\tilde{V}_g$ is the group velocity normalized by the speed of light $c$. $\tilde{V}_{\perp 0}$ is the averaged perpendicular velocity of the source population. The first order cyclotron resonance condition provides the resonance velocity,
\begin{equation}\label{Vres}
\tilde{V}_R = \chi\zeta(\omega - \Omega_{e}/\gamma) = \frac{\tilde{\omega}^2 - \sqrt{\tilde{\omega}^4 + (\tilde{\omega}^2+ \tilde{V}_p^2)(1 - \tilde{\omega}^2 -  \tilde{V}_{\perp 0}^2)}}{\tilde{\omega}^2+ \tilde{V}_p^2}\tilde{V}_p.
\end{equation} 
$\tilde{V}_R$ is dependent upon $V_{\perp0}$, because we expressed the Lorentz-factor as $\gamma = [1 - (V_R^2 + V_{\perp0}^2)/c^2]^{-1/2}$. $\tilde{V}_p = V_p/c$ is the phase velocity.

\citet{omura2011triggering} found that the frequency change of a rising-tone chorus is due to the nonlinear term $\mu_0c^2k J_B/B_w$ in the cold plasma dispersion relation. This gradual deviation in frequency can exist when the triggering wave amplitude is close to the optimum wave amplitude:

\begin{equation}\label{optampl}
 \widetilde{\Omega}_{w0}=0.81 \pi^{-5/2} \frac{Q}{\tau} \frac{s_1 \widetilde{V}_g}{s_0 \widetilde{\omega} \widetilde{U}_{t\parallel}} \left(\frac{\widetilde{\omega}_{ph} \widetilde{V}_{\perp 0} \chi}{\gamma} \right)^2 \exp\left(- \frac{\gamma^2 \widetilde{V}_{R}^2}{2 \widetilde{U}_{t\parallel}^2}\right),
\end{equation}
where $Q$ represents the depth of electron hole with typical value 0.5. $\tau = T_N / T_{tr}$ is the ratio of nonlinear transition time and nonlinear trapping period, where $T_N$ represents the time required for the formation of nonlinear current. Typical range of $\tau = 0.25-1$ is concluded from theory \cite{omura2011triggering}, simulation \cite{hikishima2012} and observation \cite{kurita2012}. $\tilde{U}_{t\parallel}= U_{t\parallel}/c$ is the parallel thermal momentum of the source population. \\
The threshold amplitude for the amplification of a chorus element is derived from the consideration that the temporal growth rate should be positive at the equator \cite{omura2009nonlinear}. Waves can only grow when the optimum amplitude is higher than the threshold amplitude and the triggering wave amplitude exceeds the threshold amplitude, 
\begin{equation}\label{treampl}
 \tilde \Omega_{th}=\frac{100\pi^3\gamma^3\xi}{\tilde\omega\tilde\omega^4_{ph}\tilde V_{\perp0}^5\chi^5} \left( \frac{\tilde a s_2 \widetilde{U}_{t\parallel}}{Q}\right)^2\exp\left(\frac{\gamma^2 \tilde V_R^2}{\widetilde{U}_{t\parallel}^2} \right),
\end{equation}
where $s_2=\frac{1}{2\xi \chi} \left\lbrace \frac{\gamma \omega}{\Omega_e} \left( \frac{V_{\perp 0}}{c}  \right)^2 - \left[ 2+\Lambda \frac{\chi^2(\Omega_e-\gamma \omega)}{\Omega_e-\omega}     \right]  \frac{V_R V_p}{c^2} \right\rbrace $ is the coefficient related to the gradient of magnetic field in the inhomogeneity ratio (Eq. (10) of \citet{omura2009nonlinear}), $\tilde a = ac^2/\Omega_{e0}=4.5c^2/(LR_E\Omega_{e0})$ is the scale length of the dipole magnetic field. $\Lambda = \omega/\Omega_e$ for inhomogeneous electron density model ($\Lambda$=1 for constant electron density model).\\
The nonlinear wave growth is:
\begin{equation}
\Gamma_N = \frac{Q \omega_{ph}^2}{2} \left(\frac{\zeta}{\Omega_w \omega}\right)^{1/2} \frac{V_g}{U_{t||}} \left(  \frac{V_{\perp 0} \chi}{c\pi\gamma} \right) exp \left( -\frac{\gamma^2 V_R^2}{2 U_{t||}^2}\right).
\end{equation}
To estimate the energetic electron density, we replace the wave amplitude in Eq. (\ref{freqsw}) with the optimum wave amplitude (\ref{optampl}):
\begin{equation}\label{hotden}
\tilde{\omega}_{ph} = \omega_{pe} \left(\frac{N_h}{N_c}\right)^{1/2} = \sqrt{\frac{\partial\omega}{\partial t} \frac{\pi^{5/2} \tau}{0.324 Q} \frac{\tilde{U}_{t||}}{\tilde{V}_g}\exp\left( \frac{\gamma^2 \tilde{V}_{R}^2}{2 \tilde{U}_{t\parallel}^2}\right)} \frac{\gamma}{\tilde{V}_{\perp 0} \chi},
\end{equation}
giving an upper-bound of $N_h$.
In the case of known thermal velocities, the number density of the source population $N_h$ can be derived from $\partial\omega/\partial t$. The relativistic linear growth-rate theory provides the estimate of $\tilde{U}_{t\parallel}$ and the average perpendicular velocity $V_{0\perp} = \sqrt{\pi/2}V_{t\perp}/c$, where we assume the bi-Maxwellian distribution.

\section{Discussion}\label{disc}

\subsection*{Case studies from EMFISIS data}

On 14 November 2012, the impact of a geomagnetic storm with a minimum Dst $\sim$ -108 nT was observable on Van Allen Probes A measurements. Chorus emissions were measured by Van Allen Probes A EMFISIS instrument from 10 to 16 UT, see Figure \ref{ANI_flux} a) . 
We have selected 16 full, strong chorus emissions from EMFISIS continuous burst mode wave data ($\mathrm{28.6 \mu s}$ time resolution and $\mathrm{\sim 12 kHz}$, maximum observable frequency  \cite{Kletzing2013}) between 11 and 12 UT. The Van Allen Probes spacecraft A was close to the plasmapause (L = 5.42 - 5.87) in the dawn sector (MLT = 4.92-5.61) and crossed the magnetic equator (mlat = 0.755 - (-0.649) $\deg$). \\
At that time, the gap at half-gyrofrequency was not formed clearly, and relatively small number of emissions existed. Therefore we concluded that a) the satellite was at the generation region and b) wave-particle interaction corresponds to the small number of emissions did not affect significantly the particle distribution of the source population .
On Figure \ref{fig1}a, three series of rising-tone emissions with large-amplitude $\mathrm{\sim 0.1-0.5 nT}$ are shown. The multicomponent wave measurement allows us to estimate the angle between the direction of propagation and the background magnetic field  $\theta$, the ellipticity and planarity of these emissions by the singular value decomposition (SVD) method  \citep{doi:10.1029/2000RS002523}. The waves exhibit quasi-parallel propagation (Fig. \ref{fig1}b), high coherence (Fig. \ref{fig1}c) and right-hand polarization (Fig. \ref{fig1}d). We present our method through the analysis of the three events in Figure \ref{inv_res}.

\begin{figure}[h]
\centering
\includegraphics[width=30pc]{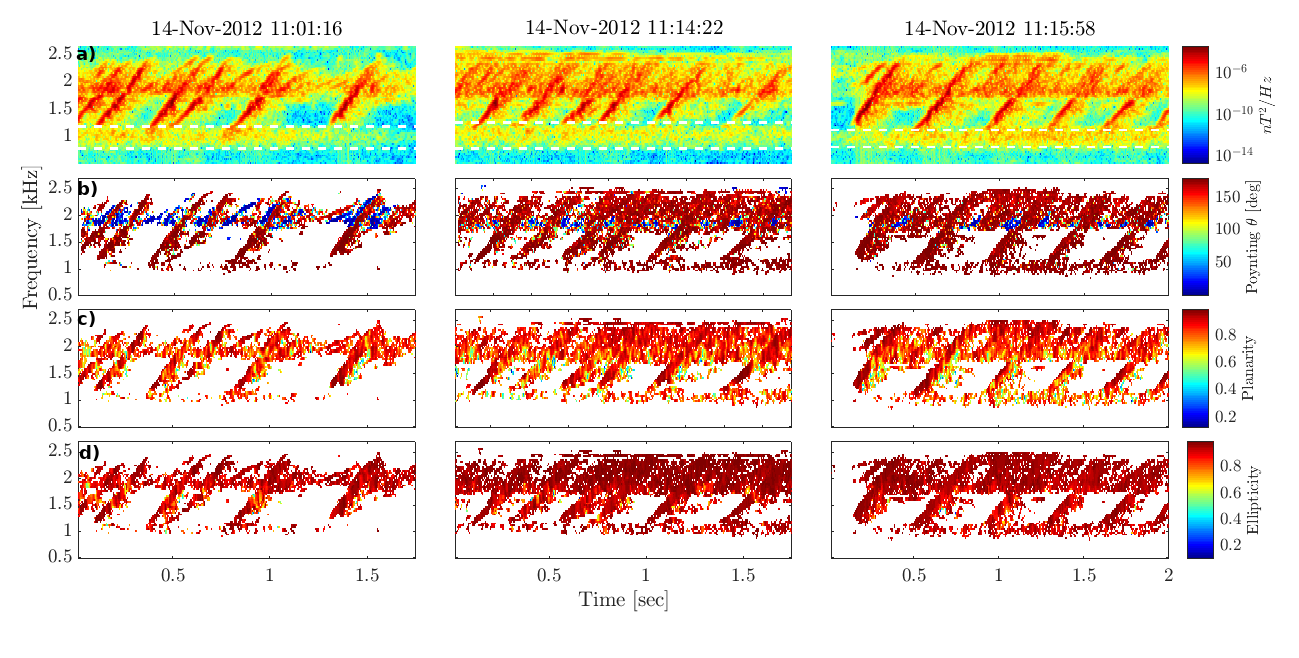}
\caption{Van Allen Probes EMFISIS-A burst data recorded on 14 November 2012: 11:01:16.67 UT (first column), 11:14:22.67 UT (second column), 11:15:58.67 UT (third column). a) Spectrogram of single-axis (BuBu) magnetic field. White dashed lines contour the assumed band of linear wave growth. b) Poynting vector angle $\theta$ with respect to the background geomagnetic field $B_0$, c) Planarity and d) Ellipticity ( magnetic PSD is greater than $\sim 10^7nT^2/Hz$)  }
\label{fig1}
\end{figure}

To estimate the parallel and the minimum perpendicular thermal velocities, first we identify the band of whistler-mode waves corresponding to the linear wave growth. The lower and upper limits of these bands are 790-1200 Hz, 790-1265 Hz and 820-1130 Hz, respectively, indicted by white dashed lines in Figure \ref{fig1}.a. From EMFISIS measurement \cite{kurth2015} we obtain $\omega_{pe}/\Omega_{e} \sim$ 5.17, 5.46 and 5.41, respectively. Assuming arbitrary $N_h$, we search for those $U_{t||}$ value, that produces the maximum linear growth rate at the mean frequency of the linear wave growth band $\omega_{rm}$. In the knowledge of $U_{t||}$, a minimum estimate for $V_{\perp0}$ can be calculated from (\ref{AC})  and
\begin{equation}
A_c =\frac{V_{t\perp}^2}{V_{t||}^2}-1 =\frac{(V_{\perp0}/\sqrt{\frac{\pi}{2}})^2}{(U_{t||}/\gamma_R)^2} - 1 
\end{equation}
In Figure \ref{inv_res}, we present the results of three emissions selected from Figure \ref{fig1}, plotting frequency sweep rates, amplitudes and growth rates. The top row of plots present the spectrogram, instantaneous frequency (blue lines) and fitted curve (dashed white lines) of rising-tone emissions.	
The relation between measured (yellow solid lines) and theoretical amplitudes are plotted in the middle panels of Figure \ref{inv_res}: optimum wave amplitudes (blue lines) are in the same order as the measured amplitudes, and have higher value than threshold amplitude (red  solid and dashed lines). Moreover, the observed amplitudes start to grow when they exceed the threshold amplitude. Although we used $\tau$= 0.25 and 0.5 for chorus-inversion, the optimum amplitude is the same, because a constant value of $\partial\omega/\partial t$ determine the product of $\tau$ and $ N_h/N_c$. Threshold amplitude is the function of $N_h/N_c$, which changes with $\tau$. Therefore using $\tau = $0.5 and 0.25 yields  a lower (red solid line) and upper (red dashed line) estimate of the threshold amplitude.  The optimum amplitude of the 2012-11-14UT11:14:24.570 event slightly differs from the measured one at higher frequencies: it can be due to an overlapping, separate upper-band chorus emissions, or convective growth. 
In the bottom row, yellow dashed lines represents $\omega_{rm}$ as they cross linear growth rate curve (dashed red lines) at the maximum. Nonlinear growth rate (blue solid ($\tau=0.5$) and dashed ($\tau=0.5$) lines)  is higher than linear growth rate, as it was proposed by \citet{summers2012a}. The frequency range of the linear  instability  is confined to $\sim 500-1500$ Hz. (We used $\tau$ = 0.25-0.5 instead of 0.25-1 in these plots, because the values of nonlinear growth rate corresponding to $\tau = 1$ are almost two order higher than linear growth rates, difficult to show the changes of the latter one in the plot.)\\
The method of chorus-inversion is sensitive to the value of $\omega_{rm}$. To obtain the standard deviation of thermal velocities, the frequency of lower and upper edge of the band are used. In Figure \ref{h_inv_res}, $V_{t||}$ (middle panel) and $V_{t\perp}$ (bottom panel) of the 16 chorus elements are marked with the squares in the middle of the red bars, and the vertical extent of the bars represent the standard deviation of $V_{t||}$ and $V_{t\perp}$. The magnitude of the standard deviation depends on the width of the linear growth rate band.
On the second step, we obtain the instantaneous frequencies of chorus emissions at the zero crossings of the wave magnetic field's perpendicular component with respect to the background magnetic field. Assuming that the frequency of the  main part of the chorus emissions is a linear function of time, we can approximate the frequency sweep rate $\partial\omega/\partial t$ with a constant value. \\
Substitute the derived values of $U_{t||}$, $V_{\perp0}$, and $\partial\omega/\partial t$ into (\ref{hotden}), $N_h/N_c$ can be calculated directly. Note, that the replacement of wave amplitude with optimum amplitude leads to an upper estimate of $N_h/N_c$. As we already mentioned, the ratio of nonlinear transition time and nonlinear trapping period $\tau$ could be between 0.25 and 1, this provides an interval for $N_h/N_c$. In the top panel of Figure \ref{h_inv_res}, $N_h/N_c$, corresponding to $\tau = $0.25-1, is shown with red errorbars and typically between 0.002 and 0.012. The red square in the middle of the errorbars corresponds to the $\tau=$0.73, which is the best fit to HOPE data (blue errorbars).
\begin{figure}[h]
\centering
\includegraphics[width=35pc]{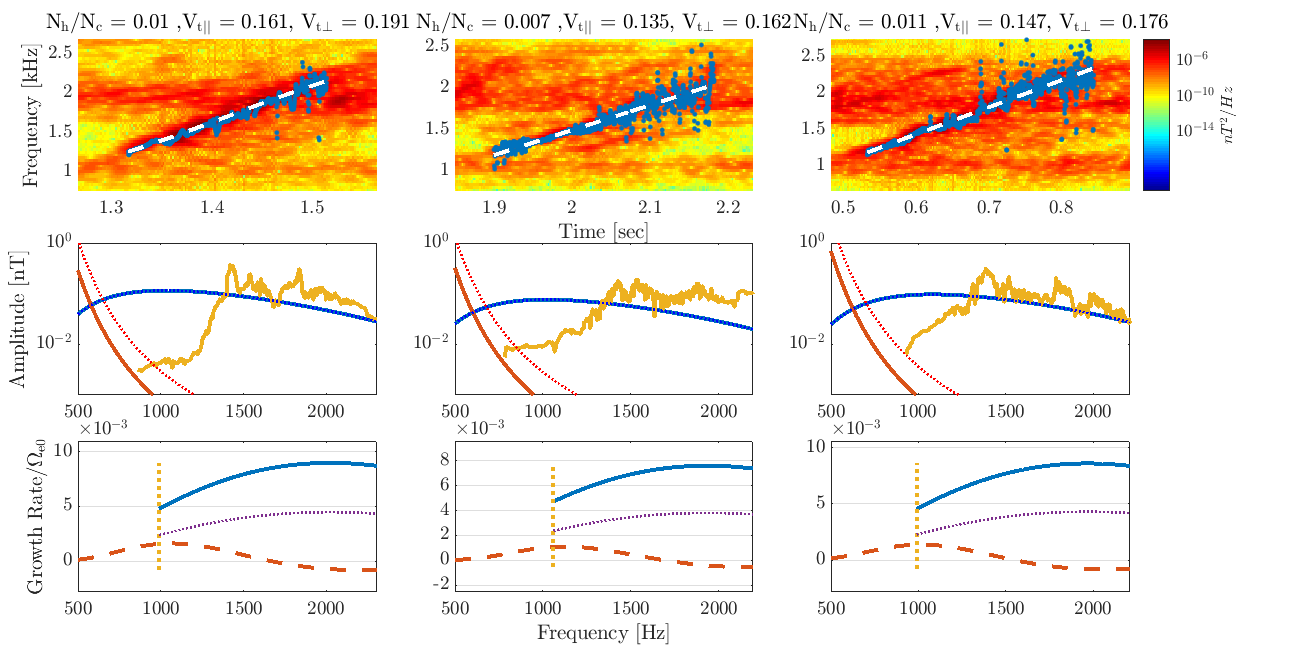}
\caption{Chorus emissions from 2012-11-14UT11:01:17.986 (left column), 2012-11-14UT11:14:24.570 (middle column) and 2012-11-14UT11:15:59.202 (right column). In the upper row the spectrogram, instantaneous frequency (blue lines) and the linear approximation (dashed white lines) of the emissions are shown. The optimum amplitudes (middle panels, blue lines) are of the same order as the measured amplitude (yellow lines), and are not affected by the change of $\tau$. Threshold amplitudes of $\tau $= 0.5 and 0.25 are plotted by solid and dotted red lines, respectively. The threshold amplitude does not depend on $\tau$ directly.  However, changes of $\tau$ modify $N_h/N_c$, which affects the threshold amplitude. Bottom panels: linear growth rate (dashed red line), $w_{rm}$ (yellow dashed lines). Nonlinear wave growth rates are plotted by blue solid ($\tau=0.5$) and dotted ($\tau=0.25$) lines. }
\label{inv_res}
\end{figure}

\subsection{Comparison of results of the inversion and in-situ measurements (HOPE data)}

The Helium, Oxygen, Proton, and Electron (HOPE) Mass Spectrometer \cite{Funsten2013} measures the fluxes of electrons and dominant ion species in the energy range of 1 eV
- 50 keV, in 36 logarithmically spaced steps (before September 2013) that was later modified to 72 log-spaced steps, at an energy resolution $\Delta E_{FWHM} /E \approx 15\%$. The $4π sr$ field of view is attained by 5 polar pixels (consisting of individual detectors) and the spin of the spacecrafts, however, HOPE data sampling is not spin synchronized. As a result, electron flux data is available as a function of energy and pitch angle. 
In this section, we compare the output of the inversion $[N_h, U_{t||}, V_{\perp0}]$, with those derived from HOPE measurements, based on the following equations (\citet{wu2013lininst} and \citet{goldstein2014VAPplas}):
  \begin{linenomath*}
  \begin{equation}
 N_h^* = 2\pi \int_0^{\pi} \int_{v_{min}}^{v_{max}} f(v,\alpha) v^2 dv  \sin \alpha d\alpha 
     = 2\pi  \sum_j \sum_i J_{ij} \left(\frac{2E_i}{m_e}\right)^{-1/2} \sin \alpha_j dE_i d\alpha_j,
\end{equation}
 \end{linenomath*}
 \begin{align}
 V_{t||}^* &= \frac{2\pi}{3N_h} \int_0^{\pi} \int_{v_{min}}^{v_{max}} v^2(\cos\alpha)^2 f(v,\alpha) v^2 dv  \sin \alpha d\alpha \nonumber\\
     &= \sqrt{\frac{2\pi^2}{m_e}} \frac{1}{3N_h} \sum_j \sum_i J_{ij} (E_i)^{1/2}  \sin \alpha_j \cos^2 \alpha_j dE_i d\alpha_j,
\end{align}
\begin{align}
 V_{t\perp}^* &= \frac{\pi}{3N_h} \int_0^{\pi} \int_{v_{min}}^{v_{max}} v^2(\sin\alpha)^2 f(v,\alpha) v^2 dv  \sin \alpha d\alpha \nonumber \\ &= \sqrt{\frac{\pi^2}{2m_e}} \frac{1}{3N_h} \sum_j \sum_i J_{ij} (E_i)^{1/2}  \sin \alpha_j \sin^2 \alpha_j dE_i d\alpha_j,
\end{align}
where $N_h^*, V_{t||}^*, V_{t\perp}^*$ are the hot electron density, parallel and perpendicular velocities from HOPE measurements. $f(v,\alpha)$ is the hot electron distribution function in the velocity $v$ and pitch angle $\alpha$ space. This theoretical description is substituted with measurable quantities, such as flux $J$, mean energy $E$ and energy width $dE_i$ of the specific energy channel. Indices $i,j$ represent the given energy channels and pitch angle bin.\\
To identify the highest and lowest energy channel of the instrument corresponding to the relativistic resonance energy of given chorus emissions, namely $[v_{min},v_{max}]$, we employ the  expression of Lorentz-factor from \citet{doi:10.1063/1.872932}:

\begin{equation}\label{lorentz}
\gamma_R = \frac{ -1 + (ck/\omega_r)[\{(ck/\omega_r)^2 -1\}(1+u_{\perp}^2/c^2)(\omega_r/\Omega_e)^2 +1]                    }{\{(ck/\omega_r)^2 -1\}(\omega_r/\Omega_e)},
\end{equation}
where $k$ is wavenumber vector, $u{\perp}$ is perpendicular momentum.
Here, we substitute the lowest and half-gyro frequency value of each chorus emissions to $\omega_{r}$, and replace $u_{t\perp}$ with the average values of perpendicular momentum derived from critical anisotropy and parallel thermal momentum. (Note, that the use of Eq.\ref{Vres} gives almost identical result.) The energy range of the comparison is based on the lower-band of the selected chorus emissions, because the upper-band of them overlaps with other upper-bands of chorus emissions that may have already modulated the lower energy part of the hot electron distribution. 

To determine the standard deviation of $N_h^*, V_{t||}^*$, and $V_{t\perp}^*$, we consider the neighboring energy channels of lowest and highest energy channels, altogether 6 channels, and we use all combinations (nine), to pick up the minimum, maximum and mean values. 
\begin{figure}[h]
\centering
\includegraphics[width=35pc]{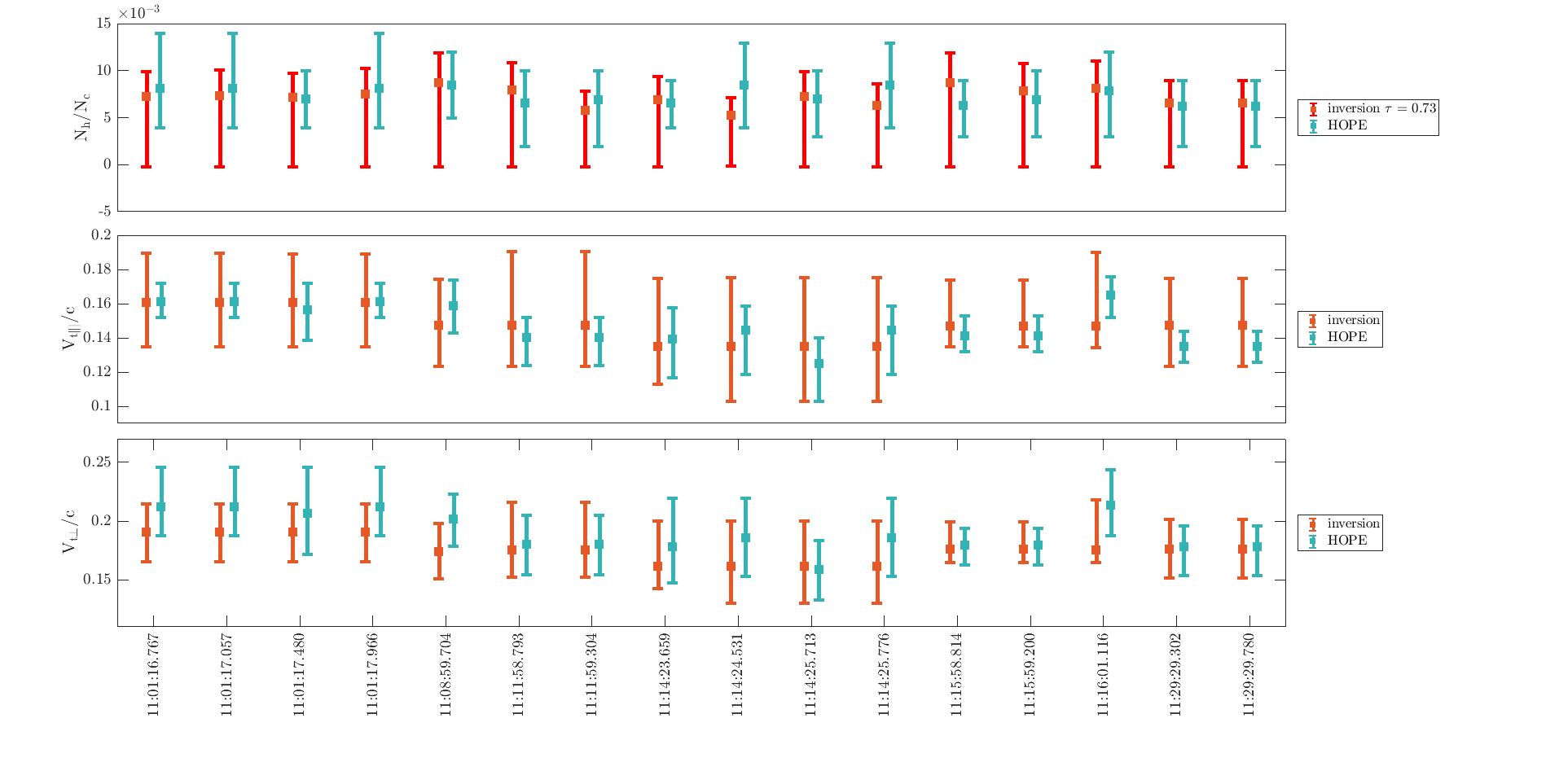}
\caption{Results of the chorus-inversion of the selected 16 chorus emissions (referred by their date). $N_h/N_c$ obtained from  HOPE measurements are shown with blue error bars. The result of the inversion for $N_h/N_c$ is a range (red errorbars), the minimum and maximum value of this interval corresponds to $\tau$ = 0.25 and 1, respectively. Top panel also shows the best fit of $\tau=$ 0.73 of the inversion to HOPE measurements with red squares.  Middle and bottom panel: parallel and  perpendicular thermal velocities from the inversion (red) and HOPE (blue) with error bars.} 

\label{h_inv_res}
\end{figure}\\
The results of ${N_h/N_c}^*, V_{t||}^*$, and $V_{t\perp}^*$ derived from HOPE measurements are plotted with light blue squares in Fig. \ref{h_inv_res}, the error bars show the standard deviation. $N_h/N_c^*$ values derived from HOPE measurements (blue) are in the range of $N_h/N_c$ (red errorbars) corresponding to $\tau$ = 0.25 and 1 . The normalized root-mean-square deviation between the HOPE ($N_h/N_c^*, V_{t||}^*$, and $V_{t\perp}^*$) and the theoretical ($N_h/N_c, V_{t||}$, and $V_{t\perp}$) values are ${N_h/N_c}_{NRMS}\sim13\%, {V_{t||}}_{NRMS} \sim6\%$, and ${V_{t\perp}}_{NRMS}\sim10\%$, respectively.%

\begin{figure}
\subfloat{\includegraphics[width = \textwidth]{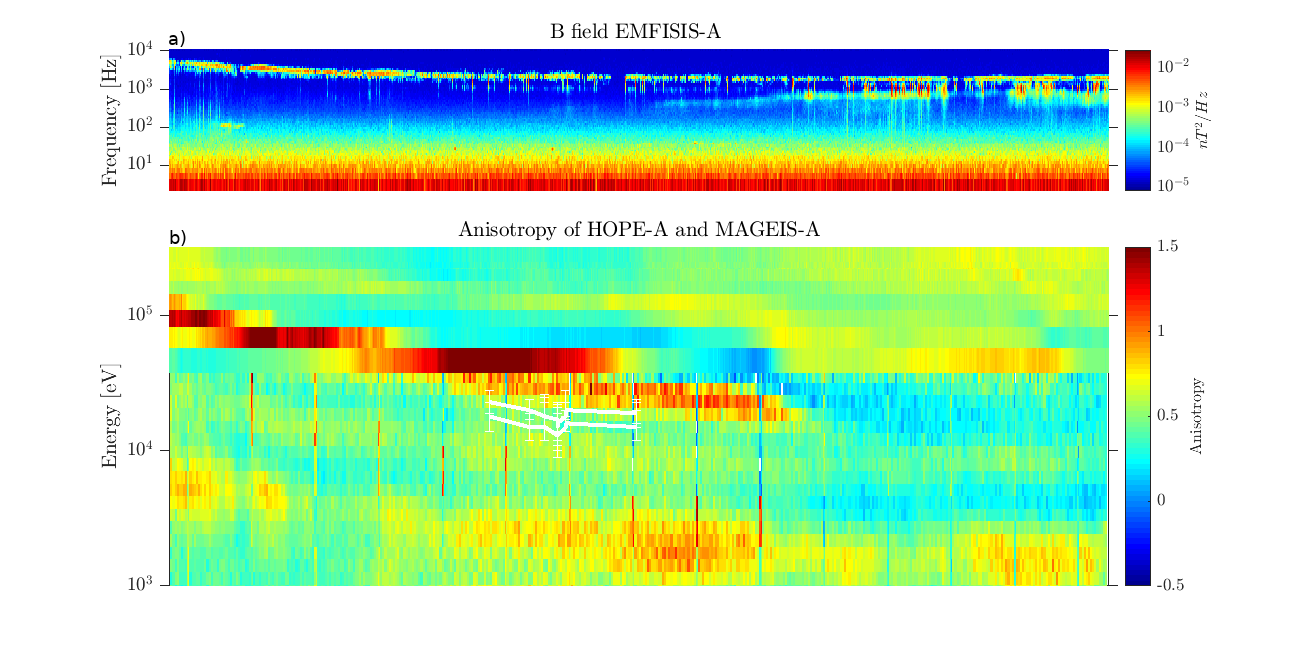}}\\ 
\subfloat{\includegraphics[width = \textwidth]{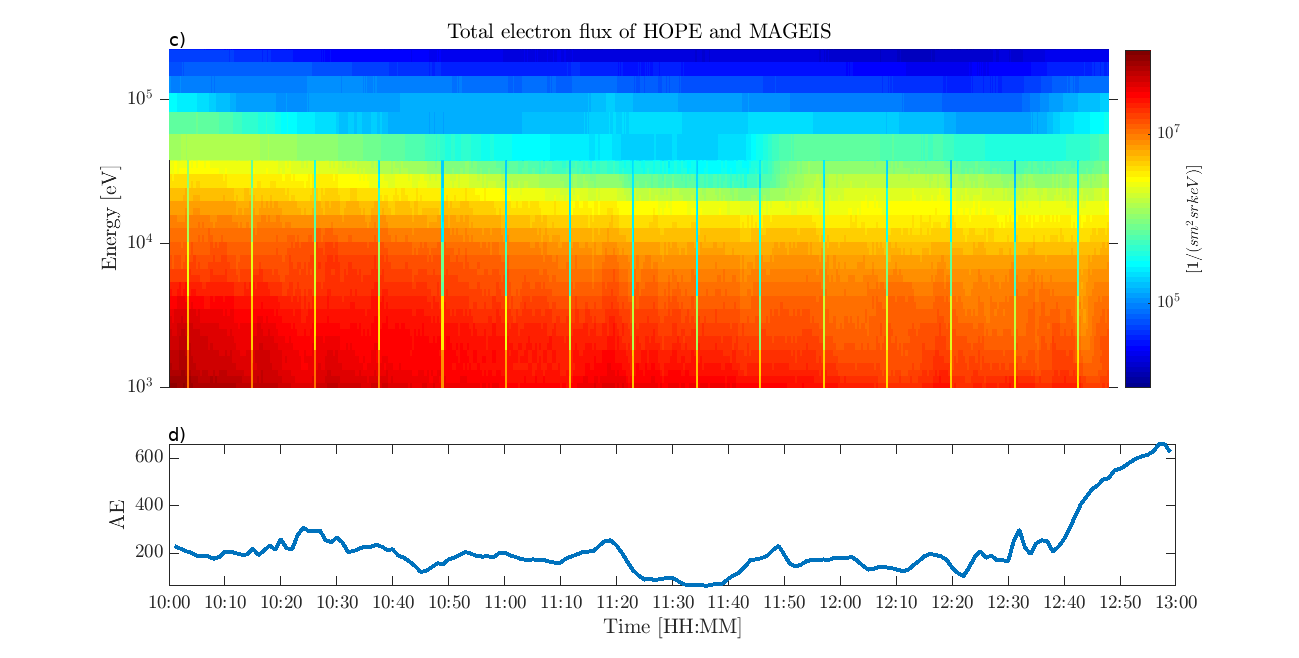}}
\caption{a. Spectrogram of single-axis (BuBu) magnetic field (top panel) measured by EMFISIS onboard Van Allen Probes spacecraft A between 10-13 UT 14 November 2012. b. Pitch-angle anisotropy map  derived from HOPE-A and MAGEIS-A measurements with resonance energy ranges of the analyzed cases (white lines). c. Total electron fluxes  measured by HOPE-A and MAGEIS-A (Top panel). d. AE index.} 
\label{ANI_flux}
\end{figure}
To affirm our results, we further analyzed  anisotropy (Fig.\ref{ANI_flux}b), omni-directional flux (Fig.\ref{ANI_flux}c), Ae index (Fig.\ref{ANI_flux}d) and wave magnetic data (Fig.\ref{ANI_flux}a) in longer timescale of 10-13UT. During this time interval, the spacecraft was flying away from the Earth to higher L shells (4-6), and was moving from the nightside to the morning sector MLT = 4-6. The pitch-angle anisotropy map in the second panel is calculated by the method of \citet{JGRA:JGRA14601} using both HOPE-A (few eV - 50keV) and MAGEIS-A (15-224 keV) particle flux measurements. The anisotropy map shows two strong anisotropic bands, one starts at 100 keV and ends $\sim 35$ keV at 10:55UT, and the other starts below 10 keV and runs parallel with the previous one. In our interpretation these anisotropic bands are the result of an injection from the plasma sheet and the plasma was accelerated in a convective transport from the nightside to the dayside.  The more isotropic region between the two bands is presumably due to wave-particle interaction between electrons in this energy range and chorus emissions (see top panel). This explanation agrees with the resonance energy ranges of the analyzed chorus emissions: the upper white line corresponds to the starting frequency of these emissions, the lower one corresponds to half the gyrofrequency.

\section{Summary and conclusion}\label{sum_con}

A new method is presented to derive $N_h, V_{t||}$, and $V_{t\perp}$ from the EMFISIS wave measurement \emph{only}. To extract these parameters from the wave data, we assumed that a) the frequency sweep rate of the chorus elements is proportional to the optimum wave amplitude, b) the optimum wave amplitude is proportional to the density of energetic electrons and c) the nonlinear wave growth generation is anticipated by linear growth rate, which is always present on the dynamic spectra as a band of whistler-mode waves close to the starting frequency of chorus emissions.
16 strong chorus emissions close to the generation region (magnetic equator) were analyzed. The output data of chorus-inversion, $N_h, V_{t||}$, and $V_{t\perp}$, were compared with the same quantities derived from the HOPE measurements in the energy range of the relativistic resonance of the selected chorus emissions, showing a good agreement (${N_h/N_c}_{NRMS}\sim13\%, {V_{t||}}_{NRMS} \sim6\%$, and ${V_{t\perp}}_{NRMS}\sim10\%$). The measured amplitudes are consistent with the optimum and threshold amplitudes of nonlinear wave growth theory, the nonlinear growth rate has positive values in the entire frequency range of chorus emissions, contrarily to the prediction of the linear growth rate theory. 
In the next step,  the method presented here will be  extended with chorus emissions recorded on the ground, replacing the in-situ wave measurements. This extension requires a suitable chorus propagation model. This way, the density of the energetic electrons can be estimated from ground data, forming a new complement or a stand-alone source of these important data (energetic electron density, parallel and perpendicular thermal velocity: $N_h, V_{t||}$, and $V_{t\perp}$ ).

\acknowledgments
The research leading to these results received funding from the Hungarian National Research, Development and Innovation Office under grant agreements NN116408 and NN116446.
This work was also supported  by JSPS KAKENHI grants 15H05815 and 17H06140.
This research was supported by the Los Alamos Space Weather Summer School, funded by the Center for Space and Earth Sciences at Los Alamos National Laboratory.
Processing and analysis of the HOPE data was supported by Energetic Particle, Composition, and Thermal Plasma (RBSP-ECT) investigation funded under NASA’s Prime contract no. NAS5-01072. All RBSP-ECT data are publicly available at the Web site http://www.RBSP-ect.lanl.gov/
All RBSP-EMFISIS data used in this paper are available from http://emfisis.physics.uiowa.edu/. The research at University of Iowa was supported under NASA prime contract NAS5-01072.

\bibliography{cikkek_kulcsszavakkal.bib}

\end{document}